# Correlation between X-ray tube current exposure time and x-ray photon number in GATE

Ignacio O. Romero, Yile Fang, and Changqing Li*
Department of Bioengineering, University of California, Merced, Merced, CA, USA.

**Abstract**

The image quality of X-ray imaging relies heavily on the X-ray output number which is dependent on the X-ray tube current and the exposure time. Hybrid X-ray imaging modalities like X-ray luminescence CT (XLCT) and X-ray fluorescence CT (XFCT) rely on the intensity of the X-ray tube to provide an accurate image reconstruction of the nanoprobe distribution in the imaging sample. A limiting factor of good image quality is the radiation dose that will be delivered to the imaging object. To accurately estimate the absorbed dose in an imaging protocol, it is better to simulate the X-ray imaging with a Monte Carlo platform such as GATE (Geant4 Application for Tomographic Emission). However, the input of GATE is a photon number of the simulated X-ray tube. So far, there is no good way to setup the photon number for a desired X-ray tube current. In this work, the accumulated radiation dose of a micro-CT X-ray tube at different current exposure times was recorded with a general-purpose ion chamber. GATE was used to model the total absorbed dose (cGy) in the sensitive volume of the ion chamber with different X-ray output numbers. A linear regression model was generated between the X-ray photon number in the GATE simulations and the tube current exposure time (mAs). The findings of this work provide an approach to correlate the X-ray tube current exposure time (mAs) to the X-ray photon number in GATE simulation of the X-ray tube.

*cli32@ucmerced.edu; http://biomedimaging.ucmerced.edu

## 1. Introduction

X-ray imaging has remained as the workhorse of medical imaging due to its fast acquisition speeds and high spatial resolution [1]. The intensity of the X-rays is dependent on the X-ray tube current and exposure time. Hybrid X-ray imaging modalities like X-ray luminescence CT (XLCT) and X-ray fluorescence CT (XFCT) rely on the intensity of the X-ray tube to provide an accurate and time efficient image of the nanoprobes in the imaging sample [2-4]. XLCT and XFCT can use X-ray excitable exogenous contrast agents to monitor drug delivery and track the progression of diseases like cancer [5-10]. However, the dose absorbed by the imaging sample from X-ray imaging is a concern and therefore it limits the image quality and applications [1, 11]. There are ways to measure the X-ray dose experimentally, but current methods are limited and/or standardized to a specific imaging object size and imaging protocol [12-14]. These methods also often include instruments that suffer from oversaturation, pile up effects, energy dependence, temperature dependence, and/or dose rate dependence which reduce the accuracy of the measurements [15-17]. Therefore, it is better to numerically simulate the X-ray imaging to accurately calculate the absorbed dose.

Among the Monte Carlo software available for medical imaging applications, the Monte Carlo software, GATE (Geant4 Application for Tomographic Emission) has gained traction in medical imaging applications [18-21]. GATE is an open-source software which was developed by the international OpenGATE collaboration as a GEANT4 wrapper that encapsulates the GEANT4 libraries specific to medical imaging and radiotherapy [22]. GATE utilizes the macro language to ease the learning curve of GEANT4 and allow GEANT4 toolkits to be more accessible to medical imaging and radiotherapy researchers [20, 22]. GATE has now allowed for the design and optimization of new medical imaging devices and radiotherapy protocols [21]. However, to model the X-ray imaging system in GATE the X-ray number is needed. So far, there is no good way to model the X-ray output number for a given X-ray tube current.

In this work, the X-ray output number from a cone beam micro-CT X-ray tube is estimated using Monte Carlo GATE. An ion chamber is used to record the accumulated exposure in Roentgens (R) of the X-ray tube for various exposure times. The exposure measurements were then converted to dose in air. The X-ray tube spectrum and measurement setup were modeled in GATE to generate a linear relationship between the radiation dose in the modeled ion chamber and the X-ray photon number. The measurement data and the trendline equations from the GATE simulations were then used to determine the X-ray output number from the dose in the experimental setup for each exposure time. Several aluminum filter thicknesses were incorporated in the experimental imaging protocol to explore the effects of filtration on X-ray output.

The paper is organized as follows. In section 2, the methods of the GATE simulations, experimental setup for source spectra measurements and dose exposure are presented. In section 3, the results showing the relationships between X-rays output and dose with different spectra are presented. The paper concludes with discussions of the results and future works.

## 2. Methods
*2.1 Experimental setup for X-ray source spectra measurement*
The spectrum of an Oxford Instruments X-ray tube (Oxford XTF5011) was acquired using an Amptek CdTe 123 spectrometer. The X-ray tube operates using the cone beam geometry and has a cone angle of 23 degrees and a tube opening of 1.143 cm diameter. To avoid oversaturation and pile up effects, a 2 mm thick, 2 mm pinhole collimator was inserted in the spectrometer and the X-ray tube was operated at 50 kVp and 0.001 mA. The spectrum was acquired for 60 seconds. The spectra of the X-ray tube with Aluminum (Al) filters of thickness 0, 0.5 mm, and 1 mm were collected and normalized.

*2.2 Experimental setup for X-ray Exposure Measurement*

The X-ray exposure was measured using an Accu-Dose system (Radcal, Monrovia, CA) with a general purpose in-beam ion chamber (10X6-6, Radcal). The Accu-Dose system meter allows for at most 4 digits when making a measurement. The ion chamber head has a diameter of 25 mm and a length of 38 mm. The active component of the ion chamber head has a volume of 6 mm$^3$. The ion chamber was positioned at the isocenter of the micro-CT imaging system which was 17.78 cm from the X-ray tube window. The X-ray tube was operated at 50 kVp and 1.0 mA. The X-ray tube system and the ion chamber were positioned inside a lead cabinet. The Accu-Dose system was placed outside the lead cabinet. The ion chamber cable was fed through a lead cabinet hole to be connected to the Accu-Dose system. The setup of the exposure acquisition and the Accu-Dose system is seen in Figure 1. The exposure was acquired in Roentgen units (R) for different exposure times: 10, 30, 60, 120, 240, and 300 seconds. Each exposure reading was repeated three times for each exposure time to reduce the variance from the Accu-Dose system measurement. The average exposure was taken from the three measurements for each exposure time. The Accu-Dose system measurements account for corrections factors due to temperature and pressure values which may be different from standard temperature and pressure in air. The Dose-to-air in the sensitive volume was found directly from the mean exposure by using the roentgen-to-cGy conversion factor, 0.876 cGy/R [11, 13, 17].

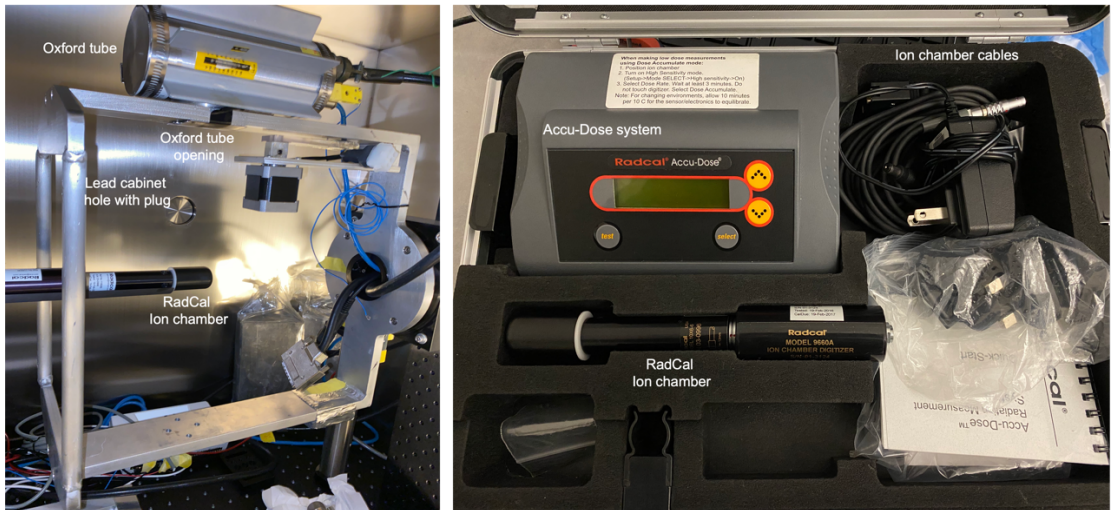

Figure 1: Experimental acquisition setup (left) and the Accu-Dose system (right). The Accu-Dose system was positioned outside of the lead cabinet for user control.

*2.3 GATE simulation setup*

A schematic and snapshot of the GATE simulation is seen in Figure 2. The ion chamber was modeled as a cylinder with wall material composed of polycarbonate. The sensitive volume of the ion chamber was also modeled as a cylinder of air with 6 cm$^3$ volume positioned at the center of the ion chamber head. The X-ray tube spectra were imported into GATE using the user spectrum function which by a fixed default normalizes the imported spectra. The GATE simulations employed the GEANT4 emstandard_opt4 physics builder to model the physical processes [20]. In the GATE simulation, the radiation dose can be stored in a 3D matrix using the

"DoseActor" tool [23]. The mass weighting algorithm was employed for the dose calculations, in which the dose was calculated as the energy deposited per unit mass of the voxel within the dose matrix. The dose matrix was discretized into 25 x 25 x 40 voxels of 1 mm$^3$ size. $10^6$, $2 \times 10^6$, $10^7$, and $2 \times 10^7$ X-rays were initialized in GATE to create a linear trendline model between the X-ray number and the radiation dose for each Al filter thickness. The absorbed dose in the modeled ion chamber sensitive volume was masked to only identify the voxels found within the sensitive volume. The dose voxels were then summed to acquire the total absorbed dose in the sensitive volume. The tube X-ray number for each measured dose was determined from a trendline model equation. All dose calculations from the modeled ion chamber sensitive volume were performed in MATLAB.

The exposure rate (R/min) was also measured using the ion chamber to validate the linear model equations for each Al filter thickness.

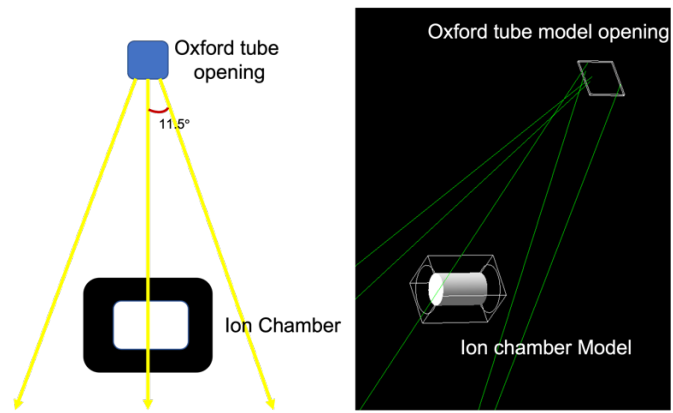

Figure 2: Schematic (left) and snapshot (right) of the GATE simulation for the dose acquisition. The sensitive volume of the ion chamber is modeled as the white cylinder.

## 3. Results

### 3.1 X-ray tube spectra

The acquired spectra for each filter case are seen in Figure 3. The L-shell energies of the tungsten anode target are clearly visible in Figure 3a when no Al filter is used. With only 0.5 mm Al thickness, the L-shell energies are effectively attenuated, and their intensity reduced as seen in Fig 3b. With 1 mm Al thickness, the L-shell energies are essentially nonexistent as seen in Figure 3c.

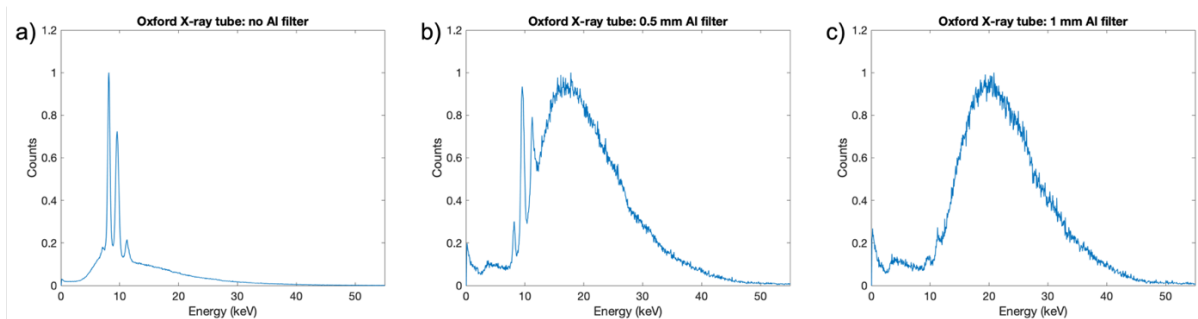

Figure 3: Spectrum of the Oxford X-ray tube with different Al thickness: a) 0 mm, b) 0.5 mm, c) 1.0 mm.

### 3.2 Measured Exposure

Figure 4 shows the linear regression plots formed from the mean of the exposure measurements (in Roentgens, R) with the ion chamber for each exposure time (milliampere-seconds, mAs) of the Oxford X-ray tube. All regression plots showed a high positive linearity with all $R^2$ values being greater than 0.99.

The highest exposure rate was seen without the Al filter due to the absorption of the L-shell energies of the tungsten anode. The plots of the 0.5 mm Al, and 1 mm Al are more similar since the L-shell energies have significantly been removed. The lowest exposure rate is seen with 1 mm Al filter thickness.

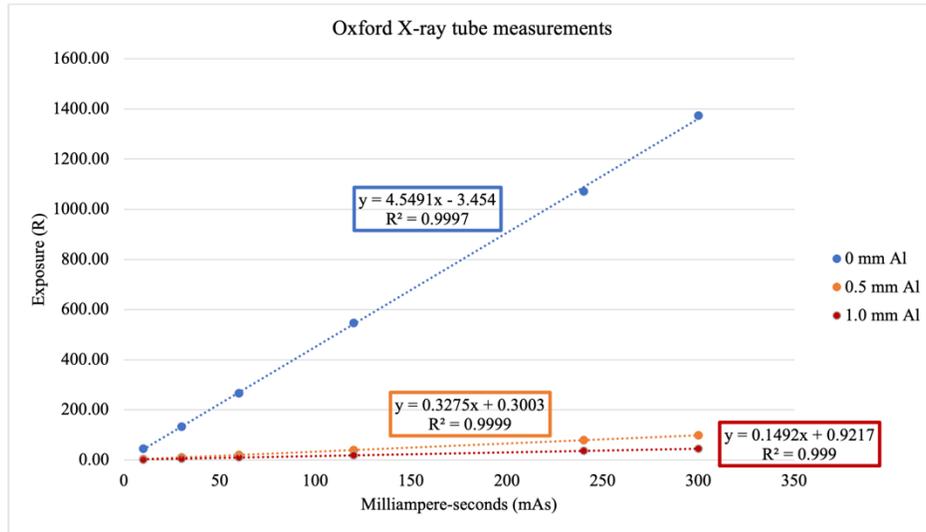

Figure 4: Linear regression plots between the mean exposure (R) and the exposure time (mAs). This plot was generated from the tube exposure measurements.

Table 1 compares and validates the accuracy of the linear trendlines equations displayed in Figure 4. Table 1 shows the percent error (PE) between the measured exposure rate and the modeled exposure rate for each Al filter thickness. The modeled exposure rates are obtained from the trendline equations in Figure 4. All modeled exposure rates are accurate within 5% of the measured exposure rate. The highest percent error was seen without the Al filter while the lowest percent error was seen with 1.0 Al filter thickness.

Table 1: Comparison of the accuracy of the modeled exposure rate to the measured exposure rate for each Al filter thickness.

| Al filter thickness (mm) | Measured Exposure Rate (R/min) | Modeled Exposure Rate (R/min) | Percent Error (%) |
| --- | --- | --- | --- |
| 0.0 | 262.0 | 272.9 | 4.16 |
| 0.5 | 19.00 | 19.65 | 3.42 |
| 1.0 | 9.082 | 8.952 | 1.43 |

*3.3 The Radiation Dose Calculated by GATE*

Figure 5 shows the linear regression plot relating the total absorbed dose in cGy and X-ray number for each Al filter thickness. The largest dose rate is observed with 0.5 mm Al filter thickness, and the lowest dose rate is observed with 1.0 mm Al filter thickness. All regression plots showed a high positive linearity with all $R^2$ values being greater than 0.999. The trendlines equations are used to convert the experimental dose readings into the X-ray output number for a given current exposure time (mAs).

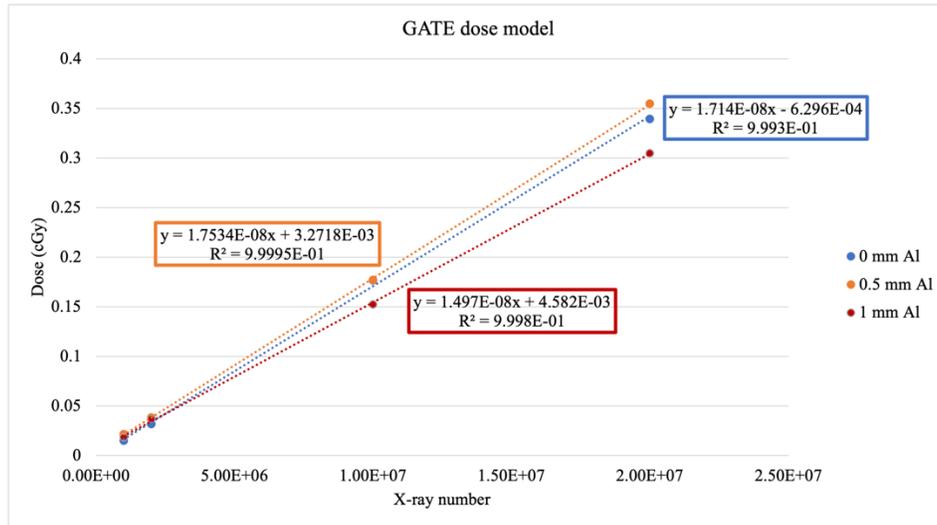

Figure 5: Linear regression plots between the total absorbed dose (cGy) and the x-ray output number modeled in GATE.

*3.4 Correlation between the Photon Number in GATE and the X-ray Tube Current*

Figure 6 shows the plots relating the X-ray number and the exposure time in milliampere-seconds (mAs) for the Oxford X-ray tube. The largest output rate is observed without the Al filter and the lowest output rate is observed with 1.0 mm Al filter thickness. All regression plots showed a high positive linearity with all $R^2$ values being greater than 0.999.

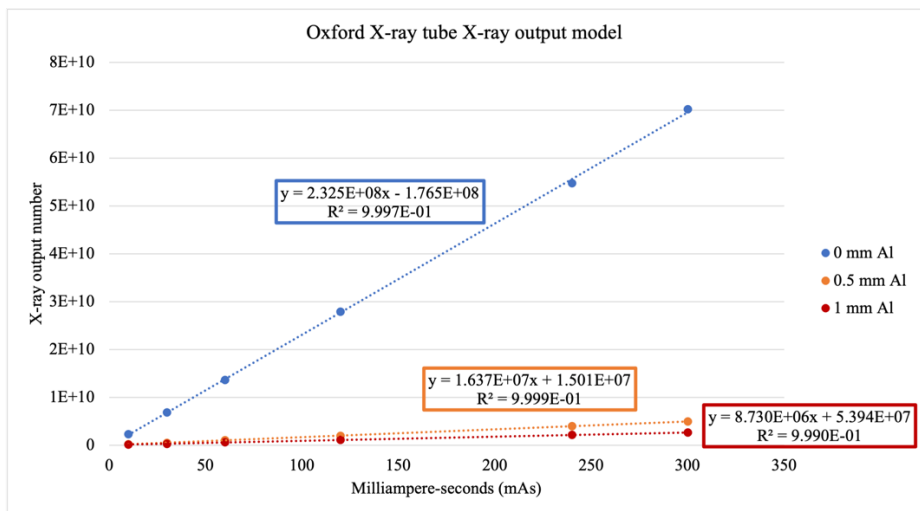

Figure 6: Linear regression plots between the X-ray output number and the exposure time (mAs).

## 4. Discussion

The accuracy of the X-ray output estimate in this work depends heavily on the specifications of the imaging system. Parameters like the beam cone angle, X-ray tube opening, source-to-detector distance (SDD), and the location of the sensitive volume inside the ion chamber head are important to include in the GATE simulation.

The measured exposure rate from Table 1 is an estimate since the Accu-Dose system meter measurement oscillates with increased exposure rates. This explains the reduction in the percentage error with increased Al filter thickness.

The increased exposure rate with decreased Al filter thickness is due to the reduced average beam energy with lower filter thickness. From Figure 3, it can be observed that the average X-ray energy increases with increases Al filter thickness since the filter removes the intense L-shell energies of the tungsten anode seen in Figure 3a. The physical process responsible for a large contribution of the dose is the photoelectric effect which is inversely proportional to the X-ray energy. Therefore, the greatest dose rate is observed without the Al filter.

In Figure 5, the greatest dose rate is observed with the 0.5 mm Al filter. This may be due to the greater effective energy with 0.5 mm Al which allows for the X-rays to become more penetrative. This leads to more dose onto the sensitive volume of the modeled ion chamber after the X-rays have traversed the ion chamber wall. However, little difference in the dose rates are observed between 0 mm Al filter and 0.5 Al filter than with the 1 mm Al filter. Without an Al filter, some of the L-shell energies of the target anode can traverse the chamber wall and deposit their energy onto the sensitive volume of the ion chamber.

In Figure 6, the X-ray output rate is greatest without the Al filter. The L-shell energies have a higher cross section than the bremsstrahlung process and therefore will produce larger amounts of radiation as observed in other literature [5, 11, 17]. However, these energies contribute mainly to the absorbed dose and require filters to be removed.

The ion chamber was positioned at the iso-center since that is the position of the imaging object and the ion chamber is fully covered by the X-ray tube cone beam. In this work, the X-ray photon number in the X-ray beam after filtering is correlated with the x-ray tube current. To correlate the photon number before filtering including the factor of x-ray self-attenuation, more detailed information of the X-ray tube product is required. However, intellectual property rights will be an obstacle to circumvent.

The trendline coefficient of determinations between X-ray number and absorbed dose is consistent with previous work and literature in which a strong linear relationship with a high coefficient of determination, $R^2 >= 0.97$ between the X-ray number and the radiation dose exists [11, 24]. However, the trendline equations are only applicable within the specified tube mAs in Figure 4 and 6. In near future works, smaller exposure times and/or tube currents will be explored to model the operational tube parameters. It is suspected that nonlinear effects in X-ray output number will be observed for millisecond exposure times.

This work can be applicable to any X-ray tube that uses the cone or fan beam geometry if the sensitive volume of the ion chamber can be completely covered by the X-ray beam. For smaller X-ray beams, like pencil beams, film dosimetry or diode edge detectors will need to be integrated into the presented method.

## 5. Conclusion

In this work, the X-ray tube output of a small animal micro-CT imaging system was estimated using Monte Carlo GATE. Exposure measurements were performed on the imaging system using an ion chamber. By comparing the simulated dose in GATE with the measured dose in the experimental setup, the X-ray photon number is found to be linearly correlated with the X-ray tube current. The linear correlation is changed with different X-ray filters. In the future work, X-ray output estimation methods will be explored in pencil beam X-ray imaging.

**Acknowledgments**
This work was funded by the NIH National Institute of Biomedical Imaging and Bioengineering (NIBIB) [R01EB026646].

**References**

[1] Huda, W. & Brad Abrahams, R. X-ray-based medical imaging and resolution. *American Journal of Roentgenology* 204, W393–W397 (2015)

[2] Romero, I. O., Fang, Y., Lun, M. & Li, C. X-ray fluorescence computed tomography (XFCT) imaging with a superfine pencil beam x-ray source. *Photonics* 8, (2021)

[3] Zhang, W., Romero, I. O. & Li, C. Time domain X-ray luminescence computed tomography: numerical simulations. *Biomedical Optics Express* 10, 372 (2019)

[4] Lun, M. C. *et al.* Focused x-ray luminescence imaging system for small animals based on a rotary gantry. *Journal of Biomedical Optics* 26,(2021)

[5] Bazalova-Carter M. The potential of L-shell X-ray fluorescence CT (XFCT) for molecular imaging. *Br J Radiol*. 2015;88(1055). doi:10.1259/bjr.20140308

[6] Bazalova M, Ahmad M, Pratx G, Xing L. L-shell x-ray fluorescence computed tomography (XFCT) imaging of Cisplatin. *Phys Med Biol*. 2014;59(1):219-232. doi:10.1088/0031-9155/59/1/219

[7] Vernekohl, D.; Xing, L. X-ray Excited Fluorescent Materials for Medical Application. *Top. Med. Chem.* 2014, *9*, 1–68

[8] T. Liu, J. Rong, P. Gao, W. Zhang and W. Liu, Cone-beam X-ray luminescence computed tomography based on X-ray absorption dosage, *J Biomed Opt* 23(2) (2018), 1–11.

[9] Lun, M. C. *et al.* Contrast agents for x-ray luminescence computed tomography. *Applied Optics* 60, 6769 (2021)

[10] D. Chen, S. Zhu, X. Chen, et al., Quantitative cone beam X-ray luminescence tomography/X-ray computed tomography imaging, *Appl Phys Lett* 105 (2014), 191104

[11] Bushberg, J. T. *The Essential physics of medical imaging*. Wolters Kluwer Health/Lippincott Williams & Wilkins, 2012; 3:229-230, 344

[12] McCollough, C. H. *et al.* CT dose index and patient dose: They are not the same thing. *Radiology* 259, 311–316 (2011)

[13] McNitt-Gray, M. F. AAPM/RSNA Physics Tutorial for Residents: Topics in CT. *RadioGraphics* 22, 1541–1553 (2002)

[14] Bazalova, M. & Verhaegen, F. Monte Carlo simulation of a computed tomography x-ray tube. *Physics in Medicine and Biology* 52, 5945–5955 (2007)

[15] Usman, S. & Patil, A. Radiation detector deadtime and pile up: A review of the status of science. *Nuclear Engineering and Technology* 50,1006–1016 (2018).

[16] Seco, J., Clasie, B. & Partridge, M. Review on the characteristics of radiation detectors for dosimetry and imaging. *Physics in Medicine and Biology* 59, R303–R347 (2014).

[17] Attix, F. H. Introduction to Radiological Physics and Radiation Dosimetry. Introduction to Radiological Physics and Radiation Dosimetry (Wiley, 1986). doi:10.1002/9783527617135

[18] Kawrakow, I. & Rogers, D. W. O. The EGSnrc Code System : Monte Carlo Simulation of Electron and Photon Transport. *System* 2001–2003 (2003)

[19] Briesmeister, J. F. *et al. MCNP-A General Monte Carlo Code for Neutron and Photon Transport, Version 3A*. *Los Alamos National Lab., NM (USA)* (1986)

[20] Agostinelli, S. *et al.* GEANT4 - A simulation toolkit. *Nuclear Instruments and Methods in Physics Research, Section A: Accelerators, Spectrometers, Detectors and Associated Equipment* 506, 250–303 (2003)

[21] Sarrut, D. *et al.* Advanced Monte Carlo simulations of emission tomography imaging systems with GATE. *Physics in Medicine and Biology* 66, (2021)

[22] S. Jan, G. Santin, D. Strul, et al., GATE: a simulation toolkit for PET and SPECT, *Phys Med Biol* 49(19) (2004), 4543–4561.

[23] D. Sarrut, M. Bardie`s, N. Boussion, et al., A review of the use and potential of the GATE Monte Carlo simulation code for radiation therapy and dosimetry applications: GATE for dosimetry, *Med Phys* 41(6) (2014), 064301.

[24] Romero, I. O. & Li, C. Radiation dose estimation for pencil beam X-ray luminescence computed tomography imaging. *Journal of X-Ray Science and Technology* **29,** 773–784 (2021